\documentclass[%
 reprint,
 amsmath,amssymb,
 aps,
]{revtex4-2}

\usepackage{graphicx}
\usepackage{dcolumn}
\usepackage{bm}
\usepackage{qcircuit}
\usepackage{subfigure}
\usepackage{MnSymbol}
\usepackage{soul}
\usepackage{comment}
\usepackage{ulem}
\usepackage{color}
\usepackage{xcolor}
\usepackage{dsfont}
\usepackage{hyperref}

\definecolor{violet}{rgb}{0.58, 0.0, 0.83}

\newcommand{\etal}{{\it et al.\/}}

\def\hh{\hskip 0.1em}
\def\sket#1{{|  \hh #1 \hh  \rangle}}
\def\sbra#1{{\langle \hh  #1 \hh |}}
\def\sbraket#1#2{{\langle \hh #1  \hh |  \hh #2 \hh  \rangle}}
\def\sexpect#1#2#3{{\langle \hh #1 \hh | \hh  #2  \hh | \hh #3 \hh \rangle}}
\def\nd{^{\vphantom{\dagger}}}
\def\ns{^{\vphantom{*}}}
\def\yd{^\dagger}
\def\CK{{\cal K}}
\def\CU{{\cal U}}
\def\CV{{\cal V}}
\def\CW{{\cal W}}
\def\tra{\textsf{Tr}\,}

\def\Re{{\rm Re}}
\def\Kc{{K_{\rm c}}}

\begin{document}

\preprint{APS/123-QED}
\title{Effects of non-integrability in a non-Hermitian time crystal}

\author{Weihua Xie}
\email{wxx180002@utdallas.edu}
\author{Michael Kolodrubetz}%
\email{mkolodru@utdallas.edu}
\affiliation{%
 Department of Physics, The University of Texas at Dallas, Richardson, Texas 75080, USA
}%


\author{Vadim Oganesyan}
\email{vadim.oganesyan@csi.cuny.edu}
\affiliation{Physics Program and Initiative for the Theoretical Sciences, The Graduate Center, CUNY, New York, New York 10016, USA}
 \affiliation{Department of Physics and Astronomy, College of Staten Island, CUNY, Staten Island, New York 10314, USA
 }

\author{Daniel P. Arovas}
\email{arovas@physics.ucsd.edu}
\affiliation{
 Department of Physics, University of California at San Diego, La Jolla, California 92093, USA
}


\date{\today}

\begin{abstract}
Time crystals are systems that spontaneously break time-translation symmetry, exhibiting repeating patterns in time. Recent work has shown that non-Hermitian Floquet systems can host a time crystalline phase with quasi-long-range order. In this work, we investigate the effect of introducing a non-integrable interaction term into this non-Hermitian time crystal model. Using a combination of numerical TEBD simulations, mean-field analysis, and perturbation theory, we find that the interaction term has two notable effects. First, it induces a shift in the phase diagram, moving the boundaries between different phases. Second, a sufficiently strong interaction induces an unexpected symmetry-breaking transition, which is not captured by the mean-field approach. Within average Hamiltonian theory, we trace this back to a ferromagnetic transition in the anisotropic non-Hermitian XXZ model. Our results demonstrate that the interplay between non-Hermitian dynamics and many-body interactions can lead to novel symmetry breaking.
\end{abstract}

\maketitle



\section{\label{sec:intro}Introduction}

Real-life quantum systems invariably interact with their environment, leading to non-unitary dynamics. These interactions are often modeled using the Lindblad formalism \cite{Manzano_2020}. When focusing on trajectories without quantum jumps, the corresponding dynamics can be described by a non-Hermitian Hamiltonian \cite{Naghiloo2019}. Non-Hermitian systems reveal distinct characteristics such as unconventional band topology \cite{RevModPhys.93.015005}, skin effects \cite{PhysRevLett.121.086803}, and exceptional points \cite{PhysRevA.86.064104, Doppler2016, Xu2016}, along with various other phenomena \cite{PhysRevA.97.032109, PhysRevLett.89.270401, Berry_2011, PhysRevLett.103.123601, PhysRevLett.104.054102}. Increasingly, experimental setups are capable of accessing non-Hermitian regimes, thus enabling the observation of these phenomena in practice \cite{Ruter2010, Li2019, doi:10.1126/science.aar7709, Gao2015, Zhang2017}. Furthermore, it was more recently found by some of us that non-Hermitian dynamics opens the door to more interesting forms of time crystals \cite{PhysRevResearch.4.013018}. More broadly, \cite{PhysRevResearch.4.013018} suggests that the canonical works by Lee and Yang, M.E. Fisher, Wu, Griffiths, and others  on statistical mechanics at complex temperature (cf.~\cite{PhysRev.87.404, PhysRev.87.410, PhysRevLett.23.17}) may be reinterpreted as non-Hermitian quantum circuits. Therefore, it seems likely that non-Hermitian Hamiltonians may have more complicated properties than Hermitian systems, which are inherited from equilibrium statistical mechanics.

One of the phases found in Basu \etal\ \cite{PhysRevResearch.4.013018} is a time crystal with quasi-long-range order in time, which differs from time crystals that are found in more conventional Hermitian systems. Time crystals -- systems that spontaneously break time-translation symmetry (TTS) to form repeating patterns in time -- were originally proposed by Wilczek for static Hamiltonians \cite{PhysRevLett.109.160401}, but such breaking of continuous time-translation symmetry was later found to be impossible \cite{PhysRevLett.114.251603}. However, the field was revived by the discovery that discrete time-translation symmetry can be broken, leading to the realization of time crystals through periodic (Floquet) driving \cite{PhysRevB.93.245146,PhysRevLett.117.090402,PhysRevLett.120.180603,PhysRevLett.120.215301,Zhang2017,Choi2017ObservationOD,Mi2022,PhysRevLett.116.250401,yousefjani2024nonhermitiandiscretetimecrystals}. The non-Hermitian time crystal from \cite{PhysRevResearch.4.013018} falls somewhere between these two classes, as it involves Floquet drive but exhibits TTS-breaking with continuously tunable period. The non-Hermitian nature of the model allows this possibility, falling outside of the considitions for the no-go theorem \cite{PhysRevLett.114.251603}.

In this work, we consider extending the non-Hermitian time crystal from Basu \etal\ \cite{PhysRevResearch.4.013018} by adding a transverse interaction between the Ising spins, which is known to break integrability. This allows us to probe the stability of the non-Hermitian time crystal, which is unclear for most driven non-Hermitian systems. On the one hand, non-Hermitian dynamics allows flow to a unique steady state, which is morally similar to a ground state and therefore is expected to be stable to weak integrability-breaking. On the other hand, the model of interest is a (non-Hermitian) Floquet system, and integrability-breaking generally leads to heating in Floquet systems which usually kills phases of matter like time crystals. Assessing this competition in the non-Hermitian time crystal is a key point of this paper.

The remainder of this paper proceeds as follows. In section \ref{s:model}, we describe the non-Hermitian Floquet-Ising model, the non-integrable interaction term, and the methods that we use to simulate its dynamics using time-evolving block decimation (TEBD). In section \ref{s:results}, we present TEBD data suggesting a shift in the phase diagram induced by interactions as well as a novel symmetry-breaking within the time-crystalline phase. We employ a mean-field treatment to analytically verify this shift and use average Hamiltonian theory to explain the additional symmetry breaking. Finally, in section \ref{s:discussion}, we conclude with a discussion of generality of these results.

\section{Model}
\label{s:model}
\begin{figure}
  \includegraphics[width=\columnwidth]{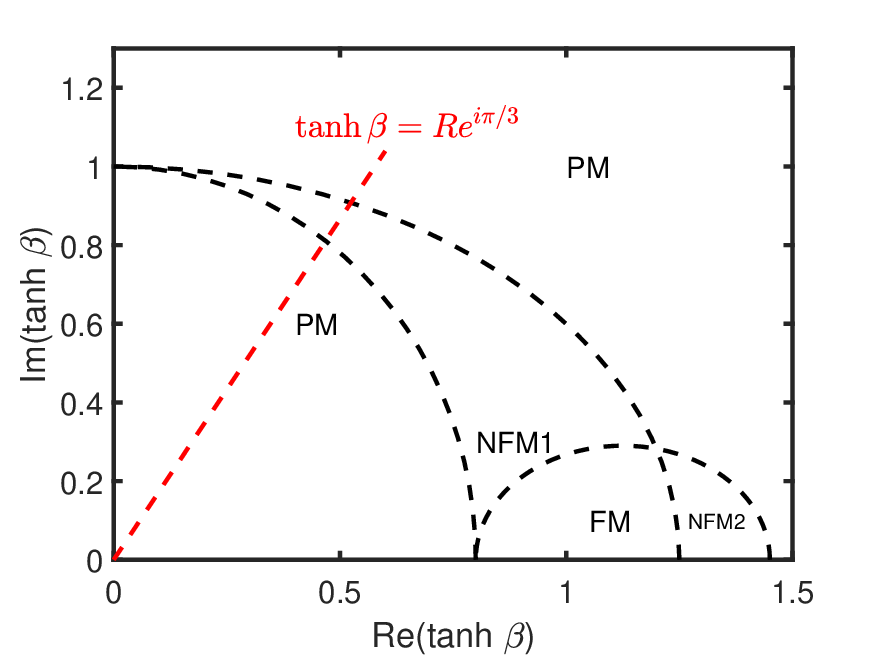}
  \caption{Phase diagram of different spatial/temporal order of our quantum circuits. PM (paramagnet): short-range temporal and spatial order. FM (ferromagnet): long-range temporal and spatial order. NFM1 (non-ferromagnet 1): quasi-long-range temporal order. NFM2 (non-ferromagnet 2): quasi-long-range spatial order. Adapted from \cite{PhysRevResearch.4.013018}.}
  \label{fig:phase_diagram}
\end{figure}

Consider the two-dimensional classical Ising model:
\begin{equation}
H = -\sum_{i,j} \big(J_x\, Z_{i,j} Z_{i+1,j} + J_{y}\, Z_{i,j} Z_{i,j+1}\big)
\end{equation}
where $Z_{i,j}$ are Pauli operators living on a square lattice. The conventional classical-to-quantum mapping takes the 2D classical Ising model at finite temperature to a one-dimensional Hermitian quantum Ising model \cite{Sachdev}. Ref.~\cite{PhysRevResearch.4.013018} considered what happens for complex temperature, which has separately studied in the theory of phase transitions as it enables zeros of the partition function, known as Lee-Yang zeros \cite{PhysRev.87.404, PhysRev.87.410}. However, \cite{PhysRevResearch.4.013018} proceeded by noting that complex temperature can also be considered as maps the classical 2D Ising model to a non-Hermitian quantum circuit, namely a non-Hermitian Floquet version of the transverse-field Ising model in which each Floquet cycle consists of first applying the bond terms and then applying the field terms:
\begin{equation}
\begin{split}
\mathcal{W} &= W_{1,2}\,W_{2,3}\,\cdots\,W_{L-1,L} = 
\exp{\left(J \sum_{j=1}^{L-1} Z_j Z_{j+1}\right)}\\
\mathcal{V} &= V_1\,V_2\,\cdots\,V_L=\exp{\left(\gamma \sum_{j=1}^L X_j \right)}
\quad,
\label{VW}
\end{split}
\end{equation}
with a non-unitary cycle evolution operator $\mathcal{U} = \mathcal{V}\mathcal{W}$.
The parameters are defined through the complex-temperature Ising mapping as 
$J = \beta J_y$ and $\tanh{\gamma} = \exp{(-2\beta J_x)}$ with parameters chosen to be anisotropic: $J_x = 1$ and 
$J_y = 0.1$. In the thermodynamic limit, $L \to \infty$, this system has multiple steady state phases (see Fig.~\ref{fig:phase_diagram}), including paramagnet, ferromagnet, and time crystal (NFM1 phase) \cite{PhysRevResearch.4.013018}. The time crystal, in particular, emerges due to a degeneracy in the decay rates of two single-particle (Majorana) modes at momenta $\pm k^\ast$. Linearizing around these points leads to a Fermi surface-like dispersion for the imaginary part of the energy, resulting in a power-law decay of 2-time correlation functions.

The time-crystalline phase is described by the two-time correlation function
\begin{equation}\label{CN}
C(N) = \langle Z_{j^\prime}(N) \, Z_j(0) \rangle =  \frac{\tra[Z_{j^\prime}\, \CU^N Z_j \, \rho_0\,
(\CU^{\dagger})^N]}{\tra[\,\CU^N \rho_0\, (\CU^{\dagger})^N]}
\end{equation}
where $N$ is number of Floquet steps and the choice of initial state $\rho_0$ is arbitrary for late-time behavior because non-Hermitian time evolution drives the system to a steady state. We therefore choose $\rho_0 \propto I$. We can also define two point spatial correlation function:
\begin{align}\label{Cy}
C(y) &= \langle Z_{{L\over 2}-y}\, Z_{{L\over 2}+y} \rangle\nd_{t \to \infty}
\end{align}
whose expectation value is evaluated in the steady state
\begin{equation}
\sket{\psi(\infty)} \propto \lim_{N \to \infty} \mathcal{U}^N \sket{\psi(0)}.
\end{equation}
In the paramagnetic phase, both spatial and temporal correlations decay exponentially, $C(N)  \sim e^{-N/\tau}$ and $C(y)  \sim e^{-y/\xi}$, where $1/\tau$ and $1/\xi$ are (generally unequal) rates of temporal and spatial decays, respectively. 
In the time-crystalline (NFM1) phase, we expect\cite{PhysRevResearch.4.013018} quasi-long-range-ordered oscillations, $C(N) \sim N^{-\alpha\nd_1}\cos{(a\nd_1 N)}$  where $a_1$ is a tunable oscillation frequency, while $C(y)\sim y^{-\alpha'_1}$. Spatial power-law decay has been  documented\cite{PhysRevResearch.4.013018} but the precise variation of time (space) exponents $\alpha_1$( $\alpha'_1$) has not been understood in detail. Similarly, for the NFM2 phase, $C(y) \sim y^{-\alpha\nd_2}\cos{(a\nd_2 y)}$. For the ferromagnetic phase, $C(N)$ and $C(y)$ converge to a non-zero values at large $N$ or $y$.

\subsection{Interaction term}
The Hermitian transverse-field Ising model is integrable. This is most readily seen by performing a Jordan-Wigner transformation to free Majorana fermions \cite{Sachdev}. For non-Hermitian models, particularly those with time-dependence, the notion of free Majoranas is more complicated; for instance, adding disorder to the field or interaction terms makes the model much harder to solve. However, as discussed in \cite{PhysRevResearch.4.013018}, the notion of free fermion integrability remains when the system is translation invariant. 

In this paper, we add an interaction term $K\sum_j X_j X_{j+1}$. In the language of Majorana fermions, 
this is a four-fermion interaction. It is known to render the system non-integrable. The Floquet operator becomes
\begin{align}
\CU_{\text{total}} &= \exp\bigg[\gamma \sum_j X_j + \stackrel{\hbox{interaction}}{\overbrace{\gamma K \sum_j
X_j  X_{j+1}  }}  \bigg] \\
&\hskip 1.4in \times \exp\bigg[\sum_j J\,Z_j Z_{j+1}\bigg]\quad.\nonumber
\end{align}

Since this model is non-integrable, we are forced to resort to approximate methods for solving it. In particular, we consider mean-field theory, exact diagonalization, and time-evolving block decimation (TEBD);  details are described below.
 
\subsection{TEBD simulations of correlation function}

The structure of non-Hermitian Floquet operators is relatively simple to implement with TEBD, as each term in $\mathcal{U}_\mathrm{total}$ can be written as a product of local matrix product operators with small bond dimensions. In order to simulate the interacting model using TEBD, we use the python package TeNPy \cite{10.21468/SciPostPhysLectNotes.5}. Non-Hermitian time evolution is obtained by decomposing each term into a Hermitian piece, which is done via conventional TEBD, and an anti-Hermitian piece, which is done via TEBD-like imaginary time evolution. A key challenge in practice is to obtain the two-time correlation function $C(N)$. We accomplished this via direct sampling of the infinite-temperature trace in terms of $z$-basis eigenstates. 

\begin{figure}
\centering
\includegraphics[width=0.5\textwidth]{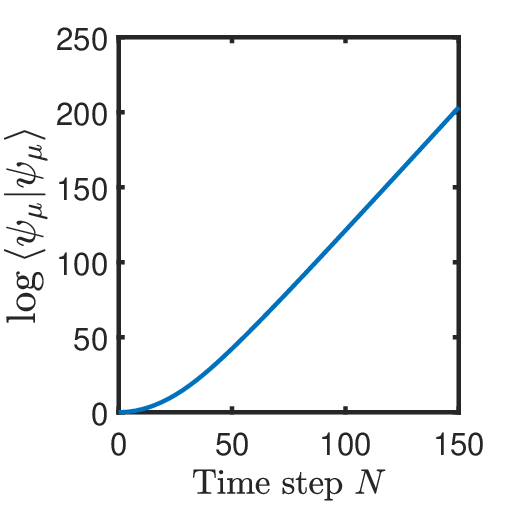}
\caption{Example fit of $\log\sbraket{\psi_\mu}{\psi_\mu}\sim kN + \text{const}$ in NFM1 phase, averaged over initial states with $\tanh{\beta} = \exp{(i\pi/3)}$.}
\label{fig:shift_fitting}
\end{figure}

To see how this works, let us rewrite the correlation function in terms of  basis states $|\psi_\mu\rangle$:
\begin{equation}
\label{MC}
C(N) = {\sum_\mu \sexpect{\psi_\mu}{Z_j\>\CU^N Z_j\, \rho_0 \,(\CU\yd)^N}{\psi_\mu} \over
\sum_\nu \sexpect{\psi_\nu}{\,\CU^N \rho_0\, (\CU\yd)^N}{\psi_\nu}}\quad.
\end{equation}
Noting that we work in the $z$-basis, $\sbra{\psi_\mu}\,Z_j = s_{\mu j}\, \sbra{\psi_\mu}$ with $s_{\mu j}=\pm 1$. Thus, 
\begin{equation}
\begin{split}
C(N) &= {\sum_\mu s_{\mu j} \sexpect{\psi_\mu}{\,\CU^N Z\ns_j\>(\CU\yd)^N}{\psi_\mu}\over
\sum_\nu\sexpect{\psi_\nu}{\,\CU^N\, (\CU\yd)^N}{\psi_\nu}}\\
&\equiv{\sum_\mu s_{\mu j}\sexpect{\psi_\mu(N)}{Z_j}{\psi_\mu(N} \over 
\sum_\nu\sbraket{\psi_\nu(N)}{\psi_\nu(N)}}
\end{split}
\end{equation}
The non-unitary evolution operator $\CU$ will make the denominator in eqn. \ref{MC} exponentially grow or decay depending on its largest magnitude eigenvalue, that is, $\sbraket{\psi(N)}{\psi(N)}\sim \exp(kN)$, which causes issues with numerical precision. To overcome this, we can fit the slope $k$ as shown in in fig. \ref{fig:shift_fitting} shift the Hamiltonians uniformly, $H \to H - k/2$. After the energy shifting, the denominator becomes  smooth. Finally, the numerator and denominator are approximated by randomly sampling the initial states $|\psi_\mu\rangle $ and keeping track of their normalization during imaginary-time evolution.

A crucial point about the efficacy of these methods is that, unlike Hermitian time-evolution, the non-Hermitian evolution here naturally prevents an unbounded growth of the entanglement. Therefore, meaningful results are able to be obtained with moderate bond dimension, even for large time delay $N$. All data shown is converged in bond dimension; some illustrations are shown in the appendix \ref{appendixB}.

\section{Results}
\label{s:results}

\begin{figure*}
  \centering
  \includegraphics[width=\textwidth]{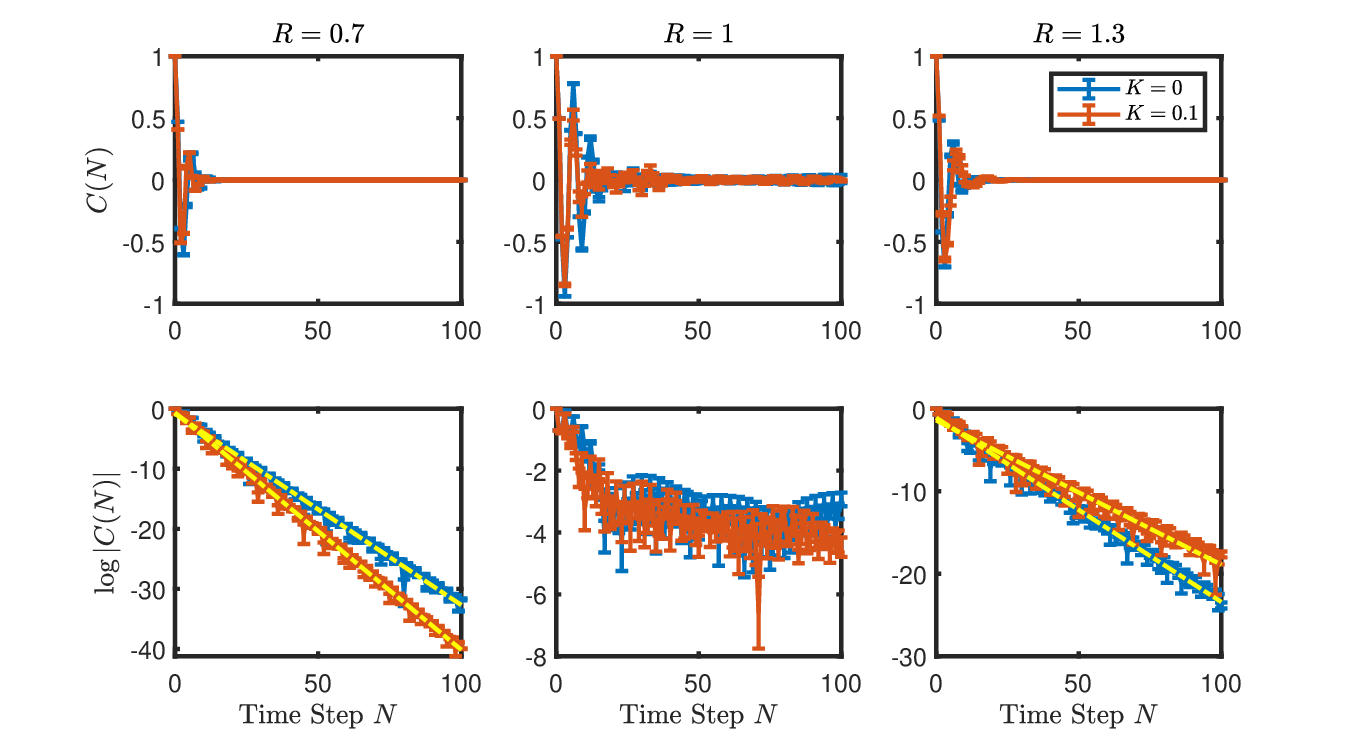}
  \caption{Upper panels are two-time correlation function, and lower panels are logarithm of the magnitude of the two-time correlation function with $\tanh{\beta} = R\,\exp(i\pi/3)$. Left panels and right panels represent PM phase, and middle panels represent NFM1 phase. The blue curve represents the data without the interaction term, while the red curve includes the interaction term. The yellow dashed lines indicate linear fits to the data. The parameters used are system size $L=150$, bond dimension $\chi=60$. \label{fig:two-time-correlations} }
\end{figure*}

The non-Hermitian Floquet-Ising model described above is simulated using TEBD; results for two-time correlations in the PM phase and NFM1 phase are shown in Figure \ref{fig:two-time-correlations}, both without integrability breaking ($K=0$) and with integrability breaking ($K=0.1$).  In order to easily tune between the phases, throughout this section we take a cut in the complex $\tanh \beta$ plane given by 
\begin{equation}
  \tanh \beta = Re^{i\pi/3} \implies R = \left| \tanh \beta \right| , 
\end{equation}
as shown in Figure \ref{fig:phase_diagram}. The point $R=1$, which lies in the middle of the NFM1 phase, is special because it corresponds to a purely real (Hermitian) magnetic field term. Tuning away from this point, the PM phase is achieved upon either increasing or decreasing $R$  by a sufficient amount.

Beginning with the PM phase, the correlations clearly decay exponentially in both cases, but the decay rate is modified by adding $K$.  Furthermore, the decay rate increases with increasing $K$ for $R=0.7$, while decreasing with increasing $K$ for $R=1.3$. This suggests that the phase transition point -- where decay rate goes to zero -- is shifting further from the $R=0.7$ point and closer to the $R=1.3$ point.

In the NFM1 phase, the conclusion is less clear. The results are consistent with long-lived oscillations with a power-law envelope, but finite size effects prevent us from definitively establishing this numerically. Note that this is a limitation of TEBD not non-integrability, as a clear power law decay for $K=0$ is also hard to establish. Therefore, in order to draw a clearer conclusion, we seek out alternative observables.


We find that a useful observable for disentangling this physics is the expectation value of the transverse magnetization $X$ in the steady state, where $X \equiv L^{-1}\sum_j X_j$. As a leading approximation, in the paramagnetic (PM) phase, $Z$-$Z$ correlations decay rapidly, suggesting that the Ising interaction is irrelevant. The dominant remaining term is a complex transverse field term, which in isolation leads to $\langle X \rangle$ values of $1$ or $-1$. In contrast, in the NFM1 phase, where the $Z$-$Z$ correlation function exhibits persistent oscillations, $\langle X \rangle$ should fall within the range $[-1,1]$.  For the case $K=0$, this behavior was seen in \cite{PhysRevResearch.4.013018}; as we will see, it is also very useful for understanding the behavior of the model with $K>0$.

\begin{figure}[!t]
\centering
\includegraphics[width=0.5\textwidth]{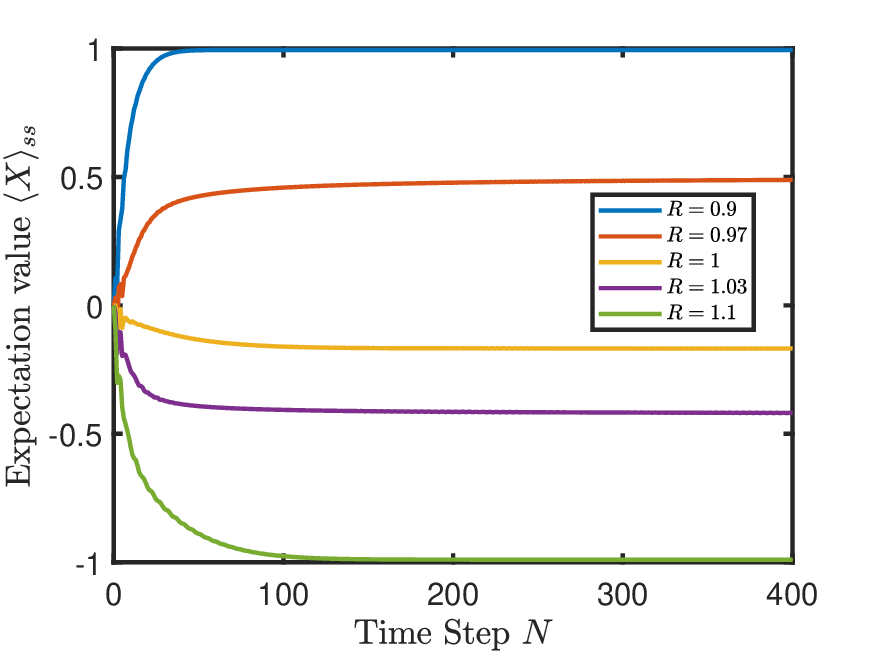}
\caption{The expectation value of $X$ in the steady states. The parameters used are system size $L=50$, $K=0.1$, and $\tanh{\beta} = R\,\exp{(i\pi/3)}$.}
\label{fig:TEBD_steady}
\end{figure}

We obtain $\langle X \rangle$ from TEBD by evolving the initial state to late times then calculating $\langle X \rangle$ in the steady state. The time evolution of $\langle X \rangle$ for a few differant values of $R$ is shown in fig. \ref{fig:TEBD_steady}. Extracting the late-time $\langle X \rangle_\mathrm{ss}$,  we see that there are kinks at the two PM to NFM1 transitions for all values of $K$ shown. 
As predicted from analyzing the two-time correlation functions, the phase boundary shifts upward to larger $R$ values when $K>0$ is added. However, we also see an unexpected new phenomenon, namely a jump in $\langle X \rangle_\mathrm{ss}$ near $R=1$. This jump is suggestive of a new first-order phase transition. We will now try to understand these two behaviors -- shifting of the critical points and the new phase transition -- using approximate treatments.

\subsection{Mean-field treatment}

As a first attempt, we notice that $K$ is relatively small, suggesting to use a mean-field decoupling of the integrability-breaking interactions. Specifically, we approximate the interaction term by
\begin{equation}
\begin{split}
\CK &= \exp\Big[K\gamma\sum_j X_j X_{j+1}\Big] \\
&\approx \exp\Bigg\{ \gamma K\sum_j X_j \! \int\limits_0^1\! dt \> \langle X(t) \rangle \Bigg\}\\
&\hskip 0.7in\times\exp\Bigg\{-\gamma K\sum_j \int\limits_0^1\! dt \> \langle X(t) \rangle^2\Bigg\}\quad.
\end{split}
\end{equation}
Furthermore, we assume that $\langle X(t) \rangle$ does not vary significantly over time, allowing us to approximate 
$\int_0^1 dt \langle X(t) \rangle = \langle X \rangle$ at the beginning of the non-unitary $\mathcal K$ gate \footnote{While it may seem that $\langle X \rangle$ is conserved during time evolution, as it commutes with the $XX$-interacting Hamiltonian, this is is not true for non-Hermitian time evolution because the different $X$-sectors can pick up different normalizations.}. Therefore, 
\begin{equation}
\CK \approx \exp\Big[2\gamma K\,\langle X\rangle \sum_j X_j \Big] \, \exp\Big[-2\gamma KL \langle\, X\rangle^2\Big]\quad.
\end{equation}
In Anderson pseudospin representation, we then solve \cite{PhysRevResearch.4.013018}
\begin{equation}
\begin{split}
\CW &= \prod_{k>0} \exp\big[2J\big(\cos(k)\,\tau_k^z  + \sin(k)\,\tau_k^y\big) \big]\\
\CV &= \prod_{k>0} A^2(\gamma) \exp{(2\gamma\tau_k^z)}\\
\CK &= \prod_{k>0} A^2(\gamma') \exp{(2\gamma'\tau_k^z)}\\
\CU &= \CK\, \CV^{1/2}\, \CW\, \CV^{1/2}  \equiv \prod_{k>0} \Theta_k
\end{split}
\end{equation}
where $\gamma' = 2\gamma K\langle X \rangle$ and $A(\gamma) = \sqrt{\coth{\gamma} - \tanh{\gamma}}$. Here we neglect the constant prefactor. Therefore,
\begin{align}
\Theta_k &= A^2(\gamma) A^2(\gamma') \exp(\gamma\tau_k^z) \exp(2\gamma'\tau_k^z)\\
&\hskip 0.35in \times\exp\big[2J\big(\cos(k)\,\tau_k^z  + \sin(k)\,\tau_k^y\big)\big] \exp(\gamma\tau_k^z)\quad.\nonumber
\end{align}

These mean-field equations are challenging to solve analytically, but can be done numerically as follows. We know that $\langle X \rangle$ lies within the interval $[-1,1]$. Therefore, we initially guess a value $\langle X \rangle_{\text{guess}}$ from this range. Next, we solve for the steady state and compute the corresponding $\langle X \rangle_{\text{actual}}$. We then iterate until these values match.

\begin{figure}
\includegraphics[width=\columnwidth]{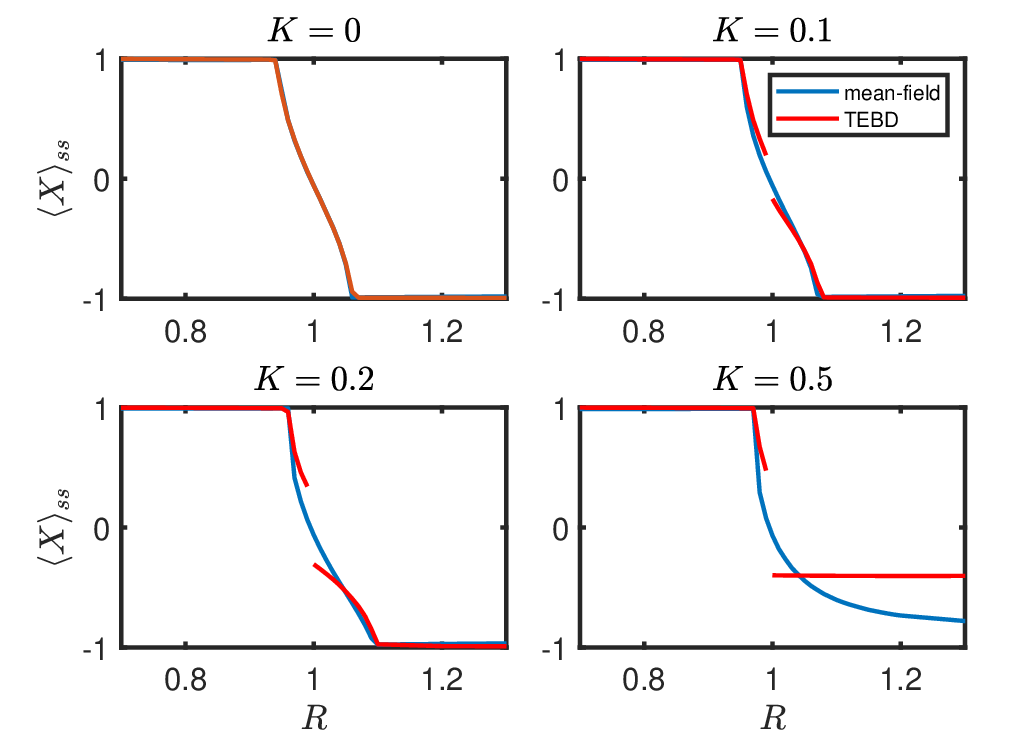}
\caption{Expectation value from mean-field treatment with different $K$ value, where $\tanh{\beta} = R\,\exp(i\pi/3)$. The panels are arranged as follows: upper left: $K = 0$, upper right: $K = 0.1$, lower left: $K = 0.2$, and lower right: $K = 0.5$. }
\label{fig:compareTEBD}
\end{figure}

The mean-field values of $\langle X \rangle_\mathrm{ss}$ are compared to the exact values in Figure \ref{fig:compareTEBD}. The shifts of the phase diagram are accurately captured by the this mean-field treatment at small and moderate $K$, meaning that this physics is just described by a self-consistent shift of the (complex) transverse field. Note that this works despite known issues with mean-field theory in one dimension because the remaining parts of the model are treated exactly via free fermions; the fluctuations missed by mean-field theory are irrelevant to shifts of the PM-NFM1 phase boundaries. However, this mean-field treatment completely misses the jump at $R=1$.

\subsection{Interaction-induced symmetry-breaking transition}

In order to understand the jump at $R=1$, we have found an average Hamiltonian treatment to be very valuable. Note first that $R=1$ corresponds to purely imaginary transverse field  $\gamma=-i\frac{\theta}{2} \equiv \gamma_0$, where $\theta = \arg \left[ \tanh \beta \right]$ is chosen to be $\pi/3$ in all numerics. Meanwhile, both $J$ and $K$ are relatively small, so we consider $J$, $K$ and $\Delta \gamma \equiv \gamma - \gamma_0$ as weak perturbations.  (Appendix \ref{appendixA}) Then, as shown in \ref{appendixA}, we can go to a rotating frame corresponding to the field term $\gamma_0$, and then time-average the remaining small terms to get the zeroth order ``Floquet Hamiltonian''
\begin{equation}
\begin{aligned}
H_{\text{eff}} &\approx i\sum_j\Big(\Delta\gamma\, X_{j}+\gamma K \, X_{j}X_{j+1}\\
&\hskip 1.0in +\frac{1}{2}J\, Z_{j}Z_{j+1}+\frac{1}{2}J\,Y_{j}Y_{j+1}\Big)\quad.\label{eq:Heff}
\end{aligned}
\end{equation}
While this is derived for $\theta$ which is a rational multiple of $\pi$, for which a finite power of $\mathcal U_0$ gives the identity (here the 6th power because $6\times (\pi/3)=2\pi$), an equivalent expression holds for the average Hamiltonian in cases with irrational angles. 

The Hamiltonian in Eq.~\ref{eq:Heff} is a non-Hermitian XXZ model with quantization axis along $x$ and a complex chemical potential term proportional to $\Delta \gamma$. For $\Delta \gamma=0$, i.e., $R=1$, one expects and XXZ model at half-filling, which has been extensively studied in both the Hermitian \cite{PhysRevB.88.245114,PhysRevB.87.035104} and non-Hermitian \cite{PhysRevB.105.205125} case. In the easy-plane limit, $|J| \gg |\gamma K|$, one expects Luttinger liquid behavior; this has been shown even for the non-Hermitian case using Bethe ansatz and bosonization \cite{PhysRevB.105.205125}. In the easy-axis limit, $|J| \ll |\gamma K|$,  one expects Ising symmetry breaking along the quantization axis $x$. Finally, for $R\neq 1$ (i.e., $\Delta \gamma \neq 0$), the chemical potential pushes the steady state away from half-filling and gives a non-symmetry-broken state that is adiabatically connected to the Luttinger liquid. In the language of our non-Hermitian Floquet model, the non-interacting case $K=0$ is precisely an easy-plane magnet. In other words, the Luttinger liquid phase in this rotating frame is equivalent to the time crystal (NFM1) phase in the lab frame. Finally, if the chemical potential strength $|\Delta \gamma|$ becomes sufficiently large, the states become either fully filled ($\langle X \rangle = 1$) or fully empty ($\langle X \rangle = -1$), which signifies a transition to the PM seen in our numerics.

\begin{figure}
\includegraphics[width=\columnwidth]{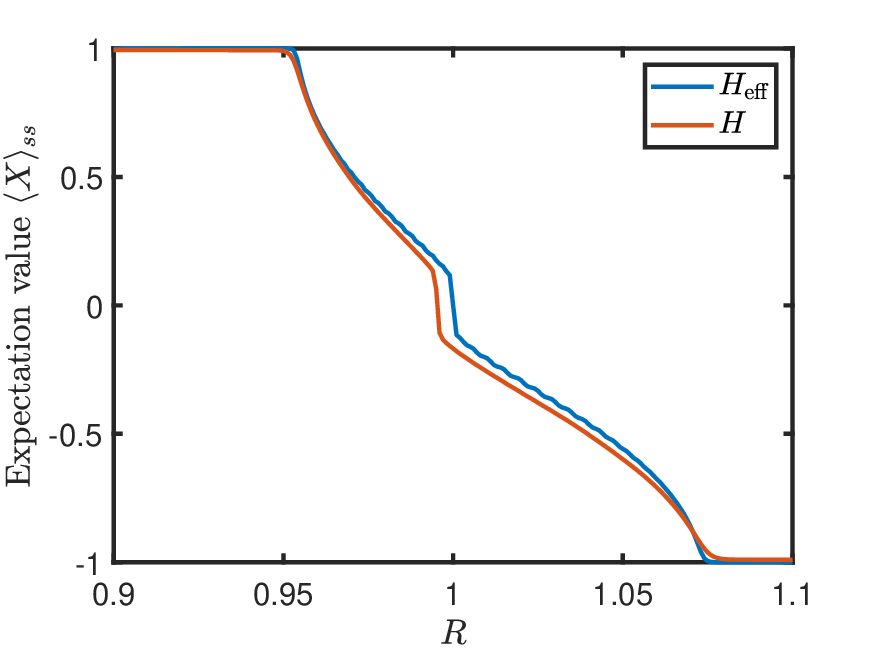}
\caption{Expectation value from $H_{\text{eff}}$ and $H$ with $\tanh \beta  = R\,\exp(i\pi/3)$ and $K = 0.1$.  }
\label{fig:L50K01}
\end{figure}

The phase structure described above is completely consistent with what is seen in the exact numerics, despite the fact that it relies on having sufficiently small $J$, $K$, and $\Delta \gamma$ that higher-order terms in the Floquet expansion can be neglected (see Sec.~\ref{appendixA} for more details). To demostrate its validity, we compare $\langle X \rangle_\mathrm{ss}$ obtained using TEBD to that of the steady state in $H_{\text{eff}}$ in Fig.~\ref{fig:L50K01}. Both clearly exhibit a jump near $R = 1$. The minor differences between the two curves can be attributed to the exclusion of higher orders of perturbation in our analysis, i.e., finite values of $J$ and $K$. 

\begin{figure}
\includegraphics[width=\columnwidth]{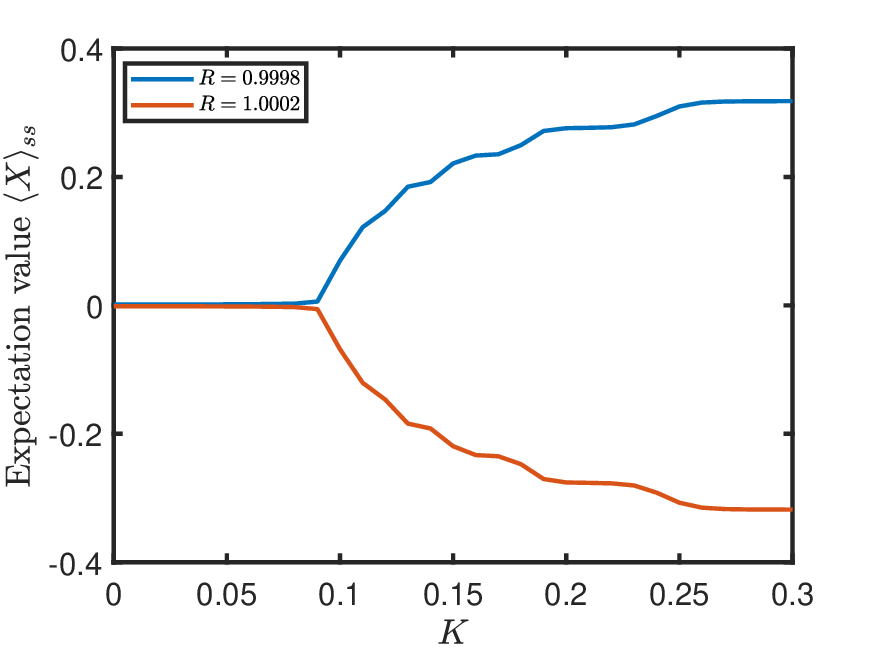}
\caption{Expectation value from $H_{\text{eff}}$. $\tanh{\beta} = \exp(i\pi/3)$. $R = 0.9998$ where $\Re(\Delta \gamma) = 0.0001$, and $R = 1.0002$ where $\Re(\Delta \gamma) = -0.0001$. We can determine the critical point as $\Kc \approx 0.085$. For $K<\Kc$, there is no observable jump; for $K>\Kc$, the magnitude of the jump increases with an increase in $K$. }
\label{fig:dgamma0}
\end{figure}

Next, to identify the critical point of this transition, we look in the limit $|\Delta \gamma|\rightarrow 0$ very near $R=1$. Specifically, we choose $R = 0.9998$ where $\Re(\Delta \gamma) = 0.0001$, and $R = 1.0002$ where $\Re(\Delta \gamma) = -0.0001$ as illustrated in fig. \ref{fig:dgamma0}. The magnetization shows a clear Luttinger liquid to $x$-ferromagnet transition at $\Kc \approx 0.085$. This confirms that the phase structure of $H_\mathrm{eff}$ is imprinted on that of the full interacting Floquet circuit.

\begin{figure}
\includegraphics[width=\columnwidth]{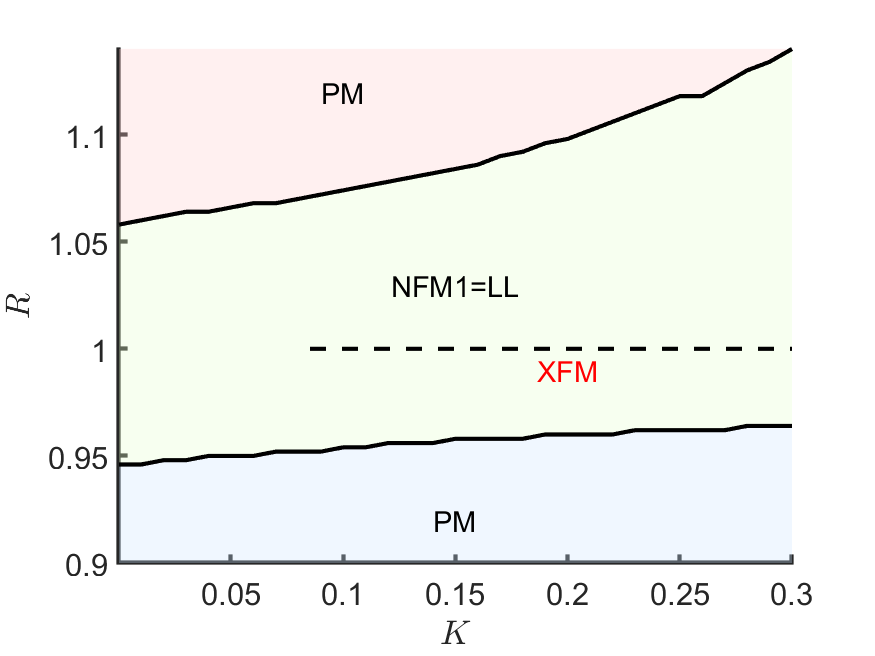}
\caption{Phase diagram after adding interaction term. $\tanh{\beta} = \exp(i\pi/3)$. Two solid black lines are the boundary between PM phase and NFM1 phase. The dashed line is the additional symmetry breaking (XFM phase). The critical point is $\Kc \approx 0.085$. For $K<\Kc$, there is no observable jump.}
\label{fig:new_phase_diagram}
\end{figure}

Combine the results from above, the phase diagram of our model as illustrated in Fig.~\ref{fig:new_phase_diagram}. Crucially, as our initial data indicated, we predict that the Luttinger liquid / NFM1 phase survives in the presence of finite $K$, even for $K > K_c$, based on the known stability of the Luttinger liquid to weak integrability breaking terms -- here given by higher-order corrections to $H_\mathrm{eff}$. Finally, we identify the new phase of matter at $R=1$ and $K > K_c$  as an $x$-ferromagnetic in the rotating frame which can be described by the easy-axis limit of the non-Hermitian XXZ model. Note that this is completely different than the $z$-ferromagnet that exists in the model for $K=0$; our integrability-breaking perturbation is vital for realizing this new phase.

\section{Discussion}
\label{s:discussion}

We introduce a non-integrable interaction term to the non-Hermitian Floquet transverse-field Ising model, calculate the correlation function, and observe that this interaction term induces a shift in the phase diagram. A mean-field treatment is employed to analytically interpret the effect of the interaction term, confirming the observed shift. Additionally, the numerical TEBD data reveals an extra symmetry breaking that is absent in the mean-field treatment. To further investigate this, we use a variant of the Floquet high-frequency expansion to derive an effective Hamiltonian, which corroborates the additional symmetry breaking as coming from a non-Hermitian XXZ Hamiltonian. These predictions are consistent with exact TEBD numerics, and similar methods are likely to allow study of other phases in this model (e.g., for $\mathrm{arg} [\tanh \beta ]  \neq \pi /3 $) and perhaps for other non-Hermitian models.

Based on the XXZ analysis, we predict that this phase structure will be quite stable, for example to the addition of other weak symmetry-respecting perturbations including those found at higher order in the Floquet expansion. This argument relies on the fact that an effective non-Hermitian Hamiltonian exists for which we expect the steady state to be quite similar to the ground state of a conventional Hermitian Hamiltonian. Therefore, for example, one should obtain conventional XXZ criticality in the vicinity of $K_c$, although the correlation functions may pick up damping factors similar to those obtained when considering complex-time correlation functions in the vicinity of a conventional quantum critical point. An intriguing open questions is whether the phase boundary should in fact remain smooth as a function of $\theta = \mathrm{arg} [\tanh \beta ]$, given that the Floquet expansion relied on a rational multiple of $\pi$ to define a finite super-Floquet cycle. One potential solution for irrational angles is to consider a nearby low-denominator rational, and then treat deviations of the angle from this rational approximation as part of $\Delta \gamma$. It remains unclear whether this analysis would give rise to a smooth phase boundary, or whether the structure of the irrational numbers would give rise to a punctured (fractal?) phase boundary as has been seen in other quasiperiodically driven systems \cite{PhysRevA.39.268}. More broadly, the unexpected symmetry breaking that we observe emphasizes the rich physics that emerges from the interplay between non-Hermitian effects and many-body interactions. These findings may have important implications for more general phase structure of driven non-Hermitian systems, an increasingly active field whose properties are increasingly able to be studied in controllable quantum systems where the desired interactions and non-Hermitian dynamics can be engineered \cite{xie2024effectnoisequantumcircuit}.

\begin{acknowledgments}
Work was performed with support from the National Science Foundation through award number DMR-1945529 (MK) and the Welch Foundation through award number AT-2036-20200401 (MK and WX).  
\end{acknowledgments}

\appendix

\begin{widetext}
\section{Effective Hamiltonian by weak perturbation}
\label{appendixA}
In this appendix, we derive the effective Hamiltonian using a Floquet high-frequency expansion. Let's start by considering Floquet perturbation theory around the point with $\tanh\beta=e^{i\theta}$ for $\theta=\pi/3$ and with $J=0$. This corresponds to $\gamma=-i\theta/2=-i\pi/6\equiv\gamma_{0}$. Then $J$, $K$, and deviations of $\gamma$ from this fine-tuned point will be treated as weak perturbations. Applying this idea directly to the period-6 time-evolution operator $\mathcal U^{6}$. Specifically, we will break it up as follows:
\begin{align}
\mathcal U & =\mathcal U_{x}\mathcal U_{zz}\\
\mathcal U_{x} & =\exp\left[\gamma\left(\sum_{j}\left(X_{j}+KX_{j}X_{j+1}\right)\right)\right]\\
\mathcal U_{zz} & =\exp\left[J\sum_{j}Z_{j}Z_{j+1}\right]
\end{align}
Define the case with $J=K=\Delta\gamma=0$ as $\mathcal U_{0}$ (where $\Delta\gamma\equiv\gamma-\gamma_{0}$).
Then first note that 
\begin{equation}
\mathcal U_{0}^{6}=e^{6\gamma_{0}\sum_{j}X_{j}}=\prod_{j}e^{-i\pi X_{j}}\propto 1.
\end{equation}
Now, to perturb around this point, we go into a rotating frame as follows:
\begin{align}
\mathcal U^{6} &=\mathcal U_{x}\mathcal U_{zz}\mathcal U_{x}\mathcal U_{zz}\mathcal U_{x}\mathcal U_{zz}\mathcal U_{x}\mathcal U_{zz}\mathcal U_{x}\mathcal U_{zz}\mathcal U_{x}\mathcal U_{zz}\\
 &=\mathcal U_{0}\mathcal U_{\Delta x}\mathcal U_{zz}\mathcal U_{0}\mathcal U_{\Delta x}\mathcal U_{zz}\mathcal U_{0}\mathcal U_{\Delta x}\mathcal U_{zz}\mathcal U_{0}\mathcal U_{\Delta x}\mathcal U_{zz}\mathcal U_{0}\mathcal U_{\Delta x}\mathcal U_{zz}\mathcal U_{0}\mathcal U_{\Delta x}\mathcal U_{zz}\\
 &=\mathcal U_{0}^{6}\mathcal U_{\Delta x}\left(\mathcal U_{0}^{\dagger}\right)^{5}\mathcal U_{zz}\mathcal U_{0}^{5}\mathcal U_{\Delta x}\left(\mathcal U_{0}^{\dagger}\right)^{4}\mathcal U_{zz}\mathcal U_{0}^{4}\mathcal U_{\Delta x}\left(\mathcal U_{0}^{\dagger}\right)^{3}\mathcal U_{zz}\mathcal U_{0}^{3}\mathcal U_{\Delta x}\left(\mathcal U_{0}^{\dagger}\right)^{2}\mathcal U_{zz}\mathcal U_{0}^{2}\mathcal U_{\Delta x}\mathcal U_{0}^{\dagger}\mathcal U_{zz}\mathcal U_{0}\mathcal U_{\Delta x}\mathcal U_{zz}
\end{align}
where we used that 
\begin{equation}
\mathcal U_{\Delta x}\equiv\exp\left[\sum_{j}\left(\Delta\gamma X_{j}+\gamma KX_{j}X_{j+1}\right)\right]
\end{equation}
commutes with $\mathcal U_{0}$. We recognize two of the terms as Heisenberg picture representations of the $ZZ$-interactions, allowing us to
rewrite as
\begin{align}
\mathcal U_{0}^{\dagger}\mathcal U_{zz}\mathcal U_{0} & =\exp\left[J\sum_{j}\mathcal U_{0}^{\dagger}Z_{j}\mathcal U_{0}\mathcal U_{0}^{\dagger}Z_{j+1}\mathcal U_{0}\right]\\
\mathcal U_{0}^{\dagger}Z\mathcal U_{0} & =e^{i\pi X/6}Ze^{-i\pi X/6}=\frac{1}{2}Z+\frac{\sqrt{3}}{2}Y\\
\left(\mathcal U_{0}^{\dagger}\right)^{2}Z\mathcal U_{0}^{2} & =-\frac{1}{2}Z+\frac{\sqrt{3}}{2}Y\\
\left(\mathcal U_{0}^{\dagger}\right)^{3}Z\mathcal U_{0}^{3} & =-Z\\
\left(\mathcal U_{0}^{\dagger}\right)^{4}Z\mathcal U_{0}^{4} & =-\frac{1}{2}Z-\frac{\sqrt{3}}{2}Y\\
\left(\mathcal U_{0}^{\dagger}\right)^{5}Z\mathcal U_{0}^{5} & =\frac{1}{2}Z-\frac{\sqrt{3}}{2}Y\\
\mathcal U_{0}^{\dagger}\mathcal U_{zz}\mathcal U_{0} & =\exp\left[\frac{J}{4}\sum_{j}\left(Z_{j}+\sqrt{3}Y_{j}\right)\left(Z_{j+1}+\sqrt{3}Y_{j+1}\right)\right]\\
 & =\exp\left[\frac{J}{4}\sum_{j}\left(Z_{j}Z_{j+1}+\sqrt{3}Y_{j}Z_{j+1}+\sqrt{3}Z_{j}Y_{j+1}+3Y_{j}Y_{j+1}\right)\right]\\
\left(\mathcal U_{0}^{\dagger}\right)^{2}\mathcal U_{zz}\mathcal U_{0}^{2} & =\exp\left[\frac{J}{4}\sum_{j}\left(Z_{j}Z_{j+1}-\sqrt{3}Y_{j}Z_{j+1}-\sqrt{3}Z_{j}Y_{j+1}+3Y_{j}Y_{j+1}\right)\right]\\
\left(\mathcal U_{0}^{\dagger}\right)^{3}\mathcal U_{zz}\mathcal U_{0}^{3} & =\exp\left[\frac{J}{4}\sum_{j}\left(4Z_{j}Z_{j+1}\right)\right]\\
\left(\mathcal U_{0}^{\dagger}\right)^{4}\mathcal U_{zz}\mathcal U_{0}^{4} & =\exp\left[\frac{J}{4}\sum_{j}\left(Z_{j}Z_{j+1}+\sqrt{3}Y_{j}Z_{j+1}+\sqrt{3}Z_{j}Y_{j+1}+3Y_{j}Y_{j+1}\right)\right]\\
\left(\mathcal U_{0}^{\dagger}\right)^{5}\mathcal U_{zz}\mathcal U_{0}^{5} & =\exp\left[\frac{J}{4}\sum_{j}\left(Z_{j}Z_{j+1}-\sqrt{3}Y_{j}Z_{j+1}-\sqrt{3}Z_{j}Y_{j+1}+3Y_{j}Y_{j+1}\right)\right]
\end{align}
Putting it together, we have
\begin{align}
\mathcal U^{6} & =\exp\left[\sum_{j}\left(\Delta\gamma X_{j}+\gamma KX_{j}X_{j+1}\right)\right]\exp\left[\frac{J}{4}\sum_{j}\left(Z_{j}Z_{j+1}-\sqrt{3}Y_{j}Z_{j+1}-\sqrt{3}Z_{j}Y_{j+1}+3Y_{j}Y_{j+1}\right)\right]\times\\
 & \;\;\;\exp\left[\sum_{j}\left(\Delta\gamma X_{j}+\gamma KX_{j}X_{j+1}\right)\right]\exp\left[\frac{J}{4}\sum_{j}\left(Z_{j}Z_{j+1}+\sqrt{3}Y_{j}Z_{j+1}+\sqrt{3}Z_{j}Y_{j+1}+3Y_{j}Y_{j+1}\right)\right]\times\\
 & \;\;\;\exp\left[\sum_{j}\left(\Delta\gamma X_{j}+\gamma KX_{j}X_{j+1}\right)\right]\exp\left[J\sum_{j}Z_{j}Z_{j+1}\right]\times\\
 & \;\;\;\exp\left[\sum_{j}\left(\Delta\gamma X_{j}+\gamma KX_{j}X_{j+1}\right)\right]\exp\left[\frac{J}{4}\sum_{j}\left(Z_{j}Z_{j+1}-\sqrt{3}Y_{j}Z_{j+1}-\sqrt{3}Z_{j}Y_{j+1}+3Y_{j}Y_{j+1}\right)\right]\times\\
 & \;\;\;\exp\left[\sum_{j}\left(\Delta\gamma X_{j}+\gamma KX_{j}X_{j+1}\right)\right]\exp\left[\frac{J}{4}\sum_{j}\left(Z_{j}Z_{j+1}+\sqrt{3}Y_{j}Z_{j+1}+\sqrt{3}Z_{j}Y_{j+1}+3Y_{j}Y_{j+1}\right)\right]\times\\
 & \;\;\;\exp\left[\sum_{j}\left(\Delta\gamma X_{j}+\gamma KX_{j}X_{j+1}\right)\right]\exp\left[J\sum_{j}Z_{j}Z_{j+1}\right]
\end{align}
Then, since there are no ``large'' terms left, we can think of this is the time evolution for time $T=6$ (in units of the original Floquet cycle) of an effective Hamiltonian, such that
\begin{equation}
\mathcal U^{6}=e^{-6iH_{\text{eff}}}\implies H_{\text{eff}}=\frac{i}{6}\ln\left(\mathcal U^{6}\right).
\end{equation}
Assuming $\left|\Delta\gamma\right|$, $\left|K\right|$, and $\left|J\right|$ are all much less than $\pi$, we can get an expression for $H_{\text{eff}}$ order-by-order using Baker-Campbell-Hausdorff. At leading order, the result is just the average Hamiltonian.
\begin{align}
H_{\text{eff}} & \approx\frac{i}{6}\left[6\sum_{j}\left(\Delta\gamma X_{j}+\gamma KX_{j}X_{j+1}\right)+3J\sum_{j}\left(Z_{j}Z_{j+1}+Y_{j}Y_{j+1}\right)\right]\\
 & =i\sum_{j}\left(\Delta\gamma X_{j}+\gamma KX_{j}X_{j+1}+\frac{1}{2}JZ_{j}Z_{j+1}+\frac{1}{2}JY_{j}Y_{j+1}\right).
\end{align}
To make the Floquet operator more symmetric, we choose $\mathcal U = \mathcal U_x^{\frac{1}{2}} \mathcal U_z \mathcal U_x^{\frac{1}{2}}$ and considering a simple case $\tanh{\beta}=\exp{(i\frac{2\pi}{3})}$, $\gamma_0 = -\frac{i\pi}{3}$.
\begin{align}
    \mathcal U^3 &= \mathcal U_x^{\frac{1}{2}} \mathcal U_z \mathcal U_x^{\frac{1}{2}} \mathcal U_x^{\frac{1}{2}} \mathcal U_z \mathcal U_x^{\frac{1}{2}} \mathcal U_x^{\frac{1}{2}} \mathcal U_z \mathcal U_x^{\frac{1}{2}}\\
    &=\mathcal U_0^{\frac{6}{2}}\mathcal U_{\Delta x}^{\frac{1}{2}}\left(\mathcal U_0^{\dagger}\right)^{\frac{5}{2}}\mathcal U_{zz}\mathcal U_0^{\frac{5}{2}}\mathcal U_{\Delta x}\left(\mathcal U_0^{\dagger}\right)^{\frac{3}{2}}\mathcal U_{zz}\mathcal U_0^{\frac{3}{2}}\mathcal U_{\Delta x}\left(\mathcal U_0^{\dagger}\right)^{\frac{1}{2}}\mathcal U_{zz}\mathcal U_0^{\frac{1}{2}}\mathcal U_{\Delta x}^{\frac{1}{2}}\\
    &= \exp\left[\frac{1}{2}\sum_{j}\left(\Delta\gamma X_{j}+\gamma KX_{j}X_{j+1}\right)\right]\exp\left[\frac{J}{4}\sum_{j}\left(Z_{j}Z_{j+1}-\sqrt{3}Y_{j}Z_{j+1}-\sqrt{3}Z_{j}Y_{j+1}+3Y_{j}Y_{j+1}\right)\right]\times\\
    & \;\;\;\exp\left[\sum_{j}\left(\Delta\gamma X_{j}+\gamma KX_{j}X_{j+1}\right)\right]\exp\left[\frac{J}{4}\sum_{j}\left(4Z_{j}Z_{j+1}\right)\right]\times\\
    & \;\;\;\exp\left[\sum_{j}\left(\Delta\gamma X_{j}+\gamma KX_{j}X_{j+1}\right)\right]\exp\left[\frac{J}{4}\sum_{j}\left(Z_{j}Z_{j+1}+\sqrt{3}Y_{j}Z_{j+1}+\sqrt{3}Z_{j}Y_{j+1}+3Y_{j}Y_{j+1}\right)\right]\times\\
    & \;\;\;\exp\left[\frac{1}{2}\sum_{j}\left(\Delta\gamma X_{j}+\gamma KX_{j}X_{j+1}\right)\right]
\end{align}
Similarly,
\begin{align}
H_{\text{eff}} & \approx\frac{i}{3}\left[3\sum_{j}\left(\Delta\gamma X_{j}+\gamma KX_{j}X_{j+1}\right)+\frac{3}{2}J\sum_{j}\left(Z_{j}Z_{j+1}+Y_{j}Y_{j+1}\right)\right]\\
 & =i\sum_{j}\left(\Delta\gamma X_{j}+\gamma KX_{j}X_{j+1}+\frac{1}{2}JZ_{j}Z_{j+1}+\frac{1}{2}JY_{j}Y_{j+1}\right).
\end{align}
\section{Convergence with bond dimension}
All data shown in this paper are converged in bond dimension, Figure \ref{fig:bond_dimension} is an example.
\label{appendixB}
\begin{figure}
\centering
\includegraphics[width=0.5\textwidth]{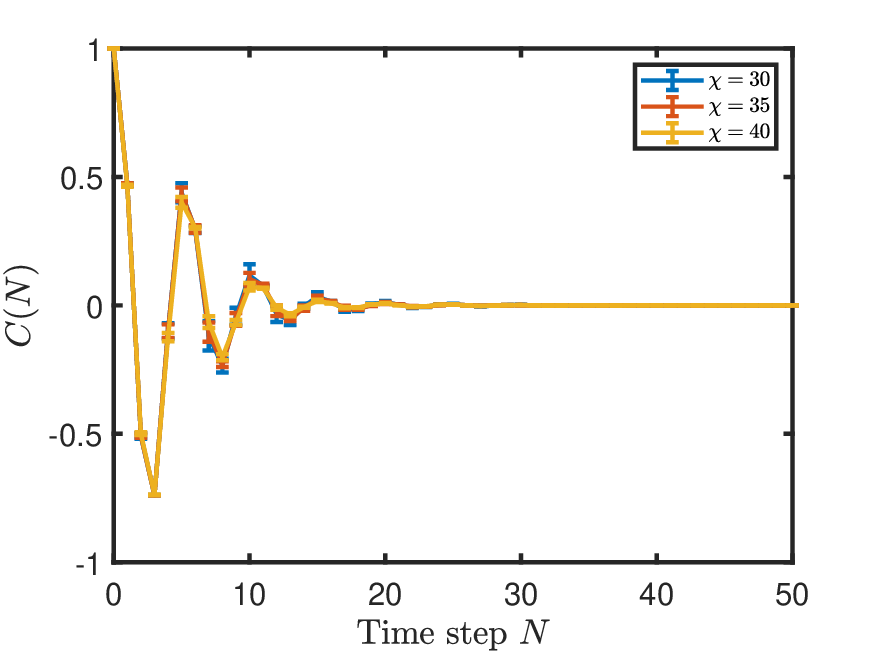}
\caption{Example of convergence with bond dimension in PM phase with $K=0.1$, $\tanh{\beta} = 0.8\exp{(i\pi/3)}$.}
\label{fig:bond_dimension}
\end{figure}
\end{widetext}


\bibliography{apssamp}

\end{document}